\documentclass[%
 reprint,
superscriptaddress,
preprintnumbers,
 amsmath,amssymb,
 aps,
]{revtex4-1}
\pdfoutput=1
\usepackage{dcolumn}

\begin{document}
\preprint{INR-TH/2015-004}
\title{Lepton flavor violating baryon decays with a single generation}

\author{A. Shkerin}
 \email{shkerin@inr.ru}
\affiliation{Institute for Nuclear Research of the Russian Academy of Sciences,
  Moscow 117312, Russia}%
\affiliation{\'{E}cole Polytechnique F\'{e}d\'{e}rale de Lausanne,
CH-1015, Lausanne, Switzerland}
\author{I. Timiryasov}
 \email{timiryasov@inr.ac.ru}
\affiliation{Institute for Nuclear Research of the Russian Academy of Sciences,
  Moscow 117312, Russia}%
\affiliation{Physics Department, Moscow State University, Vorobievy Gory,  
Moscow 119991, Russia}

\begin{abstract}
In this paper we study lepton flavor violating semileptonic decays of heavy baryons in a framework of the model \cite{Libanov:2000uf,Frere:2000dc} with large extra dimensions
and a single generation . 
Resulting branching ratios for these decays mediated by Kaluza-Klein modes of photon and Z-boson
are presented in a model-independent form.
\end{abstract}

\maketitle


\section{Introduction}

\subsection{Preliminaries}

Standard Model (SM) provides extremely accurate description of the particle physics. Therefore, it can be regarded, 
at least, as an effective low energy theory. 
To probe physics at higher energies, one can study indirect effects caused by yet undiscovered heavy 
particles predicted by various models.

Hereafter we will concentrate on the particular model, suggested in the series of works \cite{Libanov:2000uf,Frere:2000dc}, which employs
the concept of large extra dimensions \cite{Rubakov:2001kp} to solve some problems arising within the SM, such as generation replication,
fermion mass hierarchy, and gauge hierarchy. The model introduces two additional spatial dimensions forming a compact manifold (sphere). The radius of the sphere, $R$, can be understood as an effective scale of gauge fields localization \cite{Frere:2003ye}.
 6-dimensional fermions, on the other hand, are localized in a four-dimensional core of an Abrikosov-Nielsen-Olesen
 vortex \cite{Frere:2003yv}, which is formed by two auxiliary fields.
 Three zero modes of the single fermion generation in six dimensions give rise to three four-dimensional generations of the SM fermions \cite{Libanov:2000uf}.

This model can be directly probed, since it implies the existence of the 4-dimensional fermion currents that are forbidden in the SM. This fact was used in Refs. \cite{Frere:2003ye,Libanov:2012kh}, where the possible influence on K-, D- and B-meson decay physics was investigated. The experimental bounds on the branching ratios of such processes lead to the constraints on the parameters of the theory, namely, on the typical size $R$ of the localized gauge zero modes. The strongest bound found so far, which is 
\begin{equation} \label{bound}
	 \frac1R>64\mbox{~TeV},
\end{equation}
 arises from the non-observation of $K^0\to\mu e$ decay \cite{Frere:2003ye}.

Since the vast field of baryon physics remained untouched in the previous works, here we concentrate on the processes involving heavy baryons. The goal of the present paper is to calculate the partial decay widths of forbidden baryon decays.
In the framework of model with large extra dimension discussed above, the constrains on the decay widths can be translated to the bounds on $R^{-1}$.

Analytical expressions for semileptonic baryon decay rates were calculated recently, see, e.g. Ref. \cite{Garcia:1988jg}. In this work we take into account both 
weak and electromagnetic channels which makes the calculation more complicated. To obtain squared amplitudes, we use FeynCalc package \cite{Mertig:1990an}.

\subsection{Model features}
Here we describe a number of model features that are crucial for the following calculations.
Detailed description and derivation of these features 
can be found in Refs. \cite{Frere:2003ye,Frere:2000dc,Libanov:2000uf}.

The main feature is the existence of heavy Kaluza-Klein modes of the SM gauge fields. Indeed, due to periodical boundary conditions on the sphere 
of extra dimensions, 4-dimensional gauge field forms 
Kalutza-Klein tower of heavy excitations. Zero mode corresponds to the SM gauge field, and other modes
are massive and separated from the lowest mode by the mass gap $M\sim1/R$.

Higher excitations of the SM 
gauge fieldscan provide horizontal transitions between different generations of fermions \cite{Frere:2003ye}.
These transitions can be observed in Lepton Flavor Violating (LFV) processes as well as in processes with Flavor Changing Neutral Currents (FCNC).

Interaction of the fermion current $j^\mu$ with the higher excitations of the SM gauge boson is given by 4-dimensional effective Lagrangian 
\cite{Frere:2003ye}
\begin{equation}\label{GeneralForm}
\mathcal{L}_{eff}=g\cdot\text{Tr}(\textbf{A}^{\mu}\textbf{j}^*_{\mu}),
\end{equation}
where $g$ is an appropriate gauge coupling constant (i.e., $e$ for photon). Note that $\textbf{A}^{\mu}$ and $\textbf{j}_\mu$ are matrices with indices $m,n$ enumerating generation number:
\begin{equation}\label{matrix0}
\textbf{A}^{\mu}\equiv A_{m n}^\mu=\sum_{l=0}^{\infty}\left(
\begin{array}{ccc}
E_{11}^{l,0}A^{\mu}_{l,0} & E_{12}^{l,1}A^{\mu}_{l,1} & E_{13}^{l,2}A^{\mu}_{l,2} \\
E_{21}^{l,1}A^{\mu *}_{l,1} & E_{22}^{l,0}A^{\mu}_{l,0} & E_{23}^{l,1}A^{\mu}_{l,1} \\
E_{31}^{l,2}A^{\mu *}_{l,2} & E_{32}^{l,1}A^{\mu *}_{l,1} & E_{33}^{l,0}A^{\mu}_{l,0}
\end{array}\right),
\end{equation}
and
\begin{equation*}
j_{m n}^\mu=a^\dagger_m\tilde{\sigma}^\mu a_n,
\end{equation*}
where $a_n$ are two-component Weyl spinors and $E_{mn}^{l,n-m}$ are overlap factors calculated in 
\cite{Frere:2003ye},\cite{Libanov:2012kh}. Index $l$ in \eqref{matrix0} enumerates modes 
$A^{\mu}_{l,n-m}$ with four-dimensional masses $m_l^2=l(l+1)R^{-2}$. 

The situation becomes more complicated, if one also takes into account quark mixing. 
For our consideration it is important that the amplitudes of the reactions that change the total generation number $\Delta N$,
are additionally suppressed compared with those which keep it unchanged, see Ref.\cite{Frere:2003ye} for details.
%

\subsubsection*{Effective four-dimensional Lagrangian}

In what follows, we use the effective four-dimensional FCNC Lagrangian in the form
\begin{multline}
\mathcal{L}_{eff}=\mathcal{G}_{em} \mathcal{A}^\mu_{m n} \left( j^{em}_{\mu, \bar{q}_m q_n}+j^{em}_{\mu, \bar{l}_m l_n} \right)+\\
\mathcal{G}_W\mathcal{Z}^\mu_{m n} \left( j^{W}_{\mu, \bar{q}_m q_n}+j^{W}_{\mu, \bar{l}_m l_n} \right),
\label{L_eff}
\end{multline} 
where $\mathcal{A}^\mu_{m n}$ and $\mathcal{Z}^\mu_{m n}$ stand for massive neutral gauge bosons,
$\mathcal{G}_{em}$ and $\mathcal{G}_W$ are corresponding coupling constants;
$j^{em}_{\mu, \bar{q}_m q_n},j^{em}_{\mu, \bar{l}_m l_n}$ and $ j^{W}_{\mu, \bar{q}_m q_n},j^{W}_{\mu, \bar{l}_m l_n}$
are electromagnetic and weak
currents which are supposed to have the same form as the allowed within the SM currents.
Namely, for the neutral weak currents we have,
\begin{equation}\label{j}
\begin{aligned}
j^W_{\mu,\overline{l}_ml_n}=\left(-\dfrac{1}{2}+\xi\right)\overline{l}_mO^L_{\mu}l_n+\xi\overline{l}_mO^R_{\mu}l_n + h.c., \\
j^W_{\mu,\overline{q}_mq_n}=\left(T_3-\xi Q\right)\overline{q}_mO^L_{\mu}q_n-\xi Q\overline{q}_mO^R_{\mu}q_n + h.c., 
\end{aligned}
\end{equation}
\begin{equation*}
O^{R,L}_{\mu}=\gamma_{\mu}\frac{1\pm\gamma_5}{2}, 
\end{equation*}
where $\xi \equiv \sin^2\theta_W$, $T_3$ marks the component of the quark doublet and $Q$ is the charge of $q_m$ in $\vert e\vert$ unit. Electromagnetic currents read similarly,
\begin{equation}\label{jem}
\begin{aligned}
j^{em}_{\mu, \overline{l}_m l_n}=\overline{l}_m\gamma_{\mu}l_n + h.c.,\\
j^{em}_{\mu, \overline{q}_m q_n}=Q\overline{q}_m\gamma_{\mu}q_n + h.c.
\end{aligned}
\end{equation}
 Note that there is no summation over generation indices $m, n$ in \eqref{L_eff}-\eqref{jem}.

Lagrangian \eqref{L_eff} has a rather general form which makes our results applicable to any model with FCNC and LFV processes by the appropriate choice of the coupling constants
$G_e$ and $G_W$.
In the model under investigation, the matrix \eqref{matrix0} determines how heavy gauge bosons are coupled to the FCNC. From Eqs.\eqref{GeneralForm} and 
\eqref{matrix0}, we have in Eq.\eqref{L_eff},
\begin{equation}
\begin{aligned}
\mathcal{G}_{em} \mathcal{A}^\mu_{m n} &= e\sum_{l=0}^{\infty}E^{l,1}_{m n}A^{\mu}_{l,1}\\
\mathcal{G}_W \mathcal{Z}^\mu_{m n} &= \dfrac{g}{2\cos\theta_W}\sum_{l=0}^{\infty}E^{l,1}_{m n}Z^{l,1}_{\mu}
\end{aligned}
\end{equation}

\section{Baryon decays}

\subsection{General consideration}

We are interested in semileptonic baryon decays. 
According to the bounds obtained in Refs. \cite{Frere:2003ye, Libanov:2012kh}, the intermediate bosons are heavier than 
$64$ TeV. This implies that one can neglect the momentum carried by these bosons, and turn to effective four-fermion interaction. 
The amplitude of $B_i\to B_f \overline{\ell}_m \ell_n$ transition then reads,
\begin{equation}\label{Main M}
M=G_{em}H_V^{\mu} \cdot j^{em}_{\mu, \overline{l}_m l_n} +G_{W}H_{V-A}^{\mu} \cdot j^W_{\mu,\overline{l}_ml_n},
\end{equation}
with $G_{em}=\mathcal{G}_{em}/M^2$ and $G_{W}=\mathcal{G}_{W}/M^2$ being the effective four-fermion coupling constants, where the mass scale $M$ may be the same or different for electromagnetic and weak channels. 
For our model, one should also take into account the infinite number of higher excitations, that will be done in the next
sections. The hadronic current in Eq.(\ref{Main M}) can be written in the following form (\cite{Cheng:1995fe},\cite{Luo:1998wg}),
\begin{equation}\label{V-A}
H^{\mu}_{V-A}=\overline{B}_{fin}\left(F^{\mu}(q^2)+\gamma_5G^{\mu}(q^2)\right)B_{in},
\end{equation}
where
\begin{equation}\label{FormFactor}
\begin{aligned}
F^{\mu}\left(q^2\right)=f_1(q^2)\gamma^{\mu}+\frac{f_2(q^2)}{m_{B_{in}}}\sigma^{\mu\nu}q_{\nu}+\frac{f_3(q^2)}{m_{B_{in}}}q^{\mu},\\
G^{\mu}\left(q^2\right)=g_1(q^2)\gamma^{\mu}+\frac{g_2(q^2)}{m_{B_{in}}}\sigma^{\mu\nu}q_{\nu}+\frac{g_3(q^2)}{m_{B_{in}}}q^{\mu}.
\end{aligned}
\end{equation}
Here $q$ is the momentum transferred to leptons and the dimensionless quantities $f_i$, $g_i$ are the hadron form factors. To the leading order in $\vert q\vert/m_B$, we have $f_1(q^2)\approx f_1(0)$, $g_1(q^2)\approx g_1(0)$, and $f_i\approx 0$, $g_i\approx 0$, $i=2,3$. The next order approximation is $f_i(q^2)\approx f_i(0)$, $g_i(q^2)\approx g_i(0)$, $i=2,3$, and \cite{Cheng:1995fe}
\begin{equation}\label{DipoleApprox}
\begin{aligned}
f_1(q^2)\approx \frac{f_1(0)}{\left(1+q^2 / m^2_V\right)^2}\\
g_1(q^2)\approx \frac{g_1(0)}{\left(1+q^2 / m^2_A\right)^2}.
\end{aligned}
\end{equation}
Constants $m_A, m_V$ as far as the values of the form factors at zero momentum must be specify individually.
Expression for $H_V^{\mu}$ can be obtained from Eq.\eqref{V-A} by setting $G_\mu \equiv 0$.

For the decay width we have, as usual \cite{Agashe:2014kda},
\begin{equation}\label{width}
d \Gamma = \frac{1}{(2 \pi)^3} \frac{1}{32M_{B_i}^3} \vert \overline{M}\vert^2 d m_{12}^2d m_{23}^2,
\end{equation}
where $M_{B_i}$ is the mass of decaying particle, and $m_{ij}^2=(p_i+p_j)^2$, see definitions of final-state momenta in subsection \ref{SigmaDecay}.
As was already pointed out, the amplitudes with $\Delta N \neq 0$
are suppressed. We will, therefore, concentrate processes with $\Delta N = 0$. Furthermore, we restrict ourselves with the processes carried by the neutral currents described above, since we wish avoid neutrinos in a final state and detect safely all products of the reaction in order to be capable to set constraints on the flavor violation.

As we have already explained, all calculations provided below do not depend on the details of the theory whose low-energy limit we investigate, so, with obvious replacements of the coupling constants, the decay rates of the baryons will be the same in any other theory allowing such transitions. 

\subsection{An example: decays of $\Sigma$-baryons}\label{SigmaDecay}

As an example, let us perform the detailed calculations of the semileptonic decays of $\Sigma$-baryons. The triplet of strange $\Sigma$-baryons is ($\Sigma^+$, $\Sigma^-$, $\Sigma^0$).
For $\Sigma^+$ and $\Sigma^0$ there are, e.g., reactions $\Sigma^+\to pe^+\mu^-$ and $\Sigma^0\to ne^+\mu^-$, which both go through 
the neutral currents. Two other similar reactions $\Sigma^+\to pe^-\mu^+$ and $\Sigma^0\to ne^-\mu^+$ have $\vert\Delta N\vert=2$, and, hence, their decay rates are suppressed additionally by a factor $\sim 10^{-4}$, so, we do not consider them. 
Reactions with $\tau$-leptons in the final state are impossible due to kinematics.

\subsubsection*{$\Sigma^+\to pe^+\mu^-$ decay}

Both the neutral boson modes $Z^{\mu}_{l,1}$ and photon modes $A^{\mu}_{l,1}$ contribute to the amplitude of 
$\Sigma^+\to pe^+\mu^-$ decay. The part of the effective Lagrangian \eqref{L_eff}, corresponding to this process, is
\begin{multline}\label{First_vertex}
\mathcal{L}_{eff}=e\sum_{l=0}^{\infty}A^{\mu}_{l,1}E^{l,1}_{12}\left(j^{em}_{\mu, \overline{d}s}+j^{em}_{\mu, \overline{\mu}e}\right)+\\
\dfrac{g}{2\cos\theta_W}\sum_{l=0}^{\infty}E^{l,1}_{12}Z^{l,1}_{\mu}\left(j^W_{\mu,\overline{d}s}+j^W_{\mu,\overline{\mu}e}\right),
\end{multline} 
where $j^{em}_{\mu,\overline{d}s}$, $j^{em}_{\mu, \overline{\mu}e}$ and $j^W_{\mu,\overline{d}s}$, $j^W_{\mu, \overline{\mu}e}$ are the appropriate electromagnetic and weak currents. From Eqs.\eqref{j} and \eqref{jem}, we have,
\begin{equation*}
\begin{array}{cc}
j^{em}_{\mu,\overline{d}s}=-\dfrac{1}{3}\overline{d}\gamma_{\mu}s, & j^W_{\mu,\overline{d}s}=\overline{d}\gamma_{\mu}\left(-\dfrac{1}{4}+\dfrac{1}{3}\xi+\dfrac{1}{4}\gamma_5\right)s, \\
j^{em}_{\mu, \overline{\mu}e}=-\overline{\mu}\gamma_{\mu}e, & j^W_{\mu, \overline{\mu}e}=\overline{\mu}\gamma_{\mu}\left(-\dfrac{1}{4}+\xi+\dfrac{1}{4}\gamma_5\right)e.
\end{array}
\end{equation*}

In the four fermion approximation, the amplitude reads,
\begin{equation}\label{MatrixElement}
M=G_{em}H_V^{\mu} \cdot j^{em}_{\mu, \overline{l}_e l_\mu} +G_{W}H_{V-A}^{\mu} \cdot j^W_{\mu,\overline{l}_e l_\mu}.
\end{equation}
Since $p$ and $\Sigma^+$ belong to the same SU(3)-octet, we can apply the group approach to estimate the relevant form factors in 
Eqs.(\ref{FormFactor}) to the leading order in $\vert q\vert/m_{\Sigma}$ \cite{okun2013leptons}. This gives $f_1\approx 1$, $g_1\approx 1.2$.

In order to account the infinite tower of heavy excitations of the gauge bosons, we determine
the coupling constants in Eq.\eqref{MatrixElement} in the following way:
\begin{equation}\label{G}
\begin{aligned}
G_W &=\left(\dfrac{g}{2\cos\theta_W}\right)^2\sum_{l=0}^{\infty}\dfrac{\left(E^{l,1}_{12}\right)^2}{m_l^2}=\sqrt{2}G_Fm_Z^2\zeta R^2,\\
G_{em} &=e^2 \sum_{l=0}^{\infty}\dfrac{\left(E^{l,1}_{12}\right)^2}{m_l^2}=e^2\zeta R^2,
\end{aligned}
\end{equation}
where $\zeta$ represents the sum of the series \cite{Frere:2003ye}
\begin{equation*}
\zeta \equiv \sum_{l=1}^{\infty}\dfrac{\left(E^{l,1}_{12}\right)^2}{l(l+1)}\approx 0.4.
\end{equation*}

We define momenta of initial and final particles in the following way:
 $\Sigma^+ (P)\to p^+(p_1)\mu^-(p_2) e^+(p_3)$. Squared amplitude, averaged over spins of the initial baryon
 and summed over spins of the final particles \eqref{MatrixElement}, is:

\begin{widetext}
\begin{multline}\label{Msq}
\vert \overline{M}\vert^2 = 
   2 \left[m_p M_{\Sigma }  p_2\cdot p_3 \left(4 f_1^2 (1-4 \xi ) G_{\text{em}} G_W-8 f_1^2
   G_{\text{em}}^2-\left(8 \xi ^2-4 \xi +1\right) \left(f_1^2-g_1^2\right) G_W^2\right)\right)+ \\ \left(4 f_1 G_{\text{em}} G_W \left(f_1 (4 \xi -1)+g_1\right)+8 f_1^2
   G_{\text{em}}^2+G_W^2 \left(2 f_1 g_1 (4 \xi -1)+f_1^2 \left(8 \xi ^2-4 \xi +1\right)+g_1^2 \left(8 \xi ^2-4 \xi
   +1\right)\right)\right)+\\p_1\cdot p_2 P\cdot p_3 \left(2 f_1 g_1 G_W \left(-2 G_{\text{em}}-4 \xi  G_W+G_W\right)+f_1^2 \left(4
   (4 \xi -1) G_{\text{em}} G_W +\right.\right. \\ \left.\left.\left. 8 G_{\text{em}}^2+\left(8 \xi ^2-4 \xi +1\right) G_W^2\right)+g_1^2 \left(8 \xi ^2-4 \xi
   +1\right) G_W^2\right)\right].
\end{multline}
\end{widetext}
We have neglected electron mass in the last equation.
In order to compute the decay width, we rewrite Eq. \eqref{Msq} in terms of $m_{12}^2$ and $m_{23}^2$, and,
 using Eq.  \eqref{width}, we obtain numerically:
\begin{equation}\label{SigmaPWidth}
\Gamma \simeq \Gamma_W G_W^2 + \Gamma_{\text{W,em}} G_W G_{\text{em}} + \Gamma_{\text{em}} G_{\text{em}}^2,
\end{equation}
with
\begin{eqnarray*}
	\Gamma_W \simeq 6.9\cdot 10^{-8} \text{GeV}^5, \\\Gamma_{\text{W,em}} \simeq - 1.5\cdot 10^{-7} \text{GeV}^5,\\
	\Gamma_{\text{em}}  \simeq 3.2\cdot 10^{-7} \text{GeV}^5.
\end{eqnarray*}
For the value of $1/R$ consistent with the restriction \eqref{bound}, one can find for the branching ratio,
\begin{equation}\label{Br1}
	Br(\Sigma^+\to p^+\mu^- e^+)\vert_{R<R_{K\to\mu e}} < 2.4\cdot 10^{-15}.
\end{equation}

\subsubsection*{$\Sigma^0\to ne^+\mu^-$ decay}

This process is also based on the reactions $s\to d+A^{\mu}$, $s\to d+Z^{\mu}$, so it is calculated very similarly to $\Sigma^+\to pe^+\mu^-$. The only difference is in the appropriate form factors, for which we have, following to \cite{okun2013leptons}: $f_1\approx 0.7$, $g_1\approx 0.2$. 
Finally we get:
\begin{eqnarray*}
	\Gamma_W \simeq 1.2\cdot 10^{-8} \text{GeV}^5, \\\Gamma_{\text{W,em}} \simeq - 2.2\cdot 10^{-8} \text{GeV}^5,\\
	\Gamma_{\text{em}}  \simeq 1.6\cdot 10^{-7} \text{GeV}^5.
\end{eqnarray*}
For the value of $1/R$ consistent with the restriction \eqref{bound}, one can find for the branching ratio:
\begin{equation}\label{Br2}
	\text{Br}(\Sigma^0\to n\mu^- e^+)\vert_{R<R_{K\to\mu e}} < 1.3\cdot 10^{-24}.
\end{equation}
The reason for the sharp difference between the low limit restrictions \eqref{Br1} and \eqref{Br2} is in the distinction between mean lifetimes of the baryons: $\tau_{\Sigma^+}\approx 0.8\cdot 10^{-10} s$ and $\tau_{\Sigma^0}\approx 7.4\cdot 10^{-20} s$. Due to this difference, the reaction $\Sigma^+\to pe^+\mu^-$ is more preferable for searches in experiments.

\subsection{Charmed and beauty baryons}

\subsubsection*{Charmed baryons}

Among the charmed baryons, the best restrictions on the forbidden semileptonic decays so far have been obtained for $\Lambda^+_c$-baryon. To avoid the additional suppression, we will consider $e\to\mu$ lepton current transition, since $c\to u$ transition between the first and the second generations of quarks have $\vert\Delta N\vert=1$. The interaction vertex for these reactions has the following form,
\begin{multline}\label{SecondVertex}
\mathcal{L}_{eff}=e\sum_{l=0}^{\infty}A^{\mu}_{l,1}E^{l,1}_{12}\left(j^{em}_{\mu, \overline{u}c}+j^{em}_{\mu, \overline{\mu}e}\right)+\\
\dfrac{g}{2\cos\theta_W}\sum_{l=0}^{\infty}E^{l,1}_{12}Z^{l,1}_{\mu}\left(j^W_{\mu,\overline{u}c}+j^W_{\mu,\overline{\mu}e}\right),
\end{multline} 
with
\begin{equation*}
\begin{array}{cc}
j^{em}_{\mu,\overline{u}c}=\dfrac{2}{3}\overline{d}\gamma_{\mu}s, & j^W_{\mu,\overline{u}c}=\overline{u}\gamma_{\mu}\left(\dfrac{1}{4}-\dfrac{2}{3}\xi-\dfrac{1}{4}\gamma_5\right)c, \\
j^{em}_{\mu, \overline{\mu}e}=-\overline{\mu}\gamma_{\mu}e, & j^W_{\mu, \overline{\mu}e}=\overline{\mu}\gamma_{\mu}\left(-\dfrac{1}{4}+\xi+\dfrac{1}{4}\gamma_5\right)e.
\end{array}
\end{equation*}
This leads to the following transitions:
\begin{eqnarray}\label{charmed reactions}
\Lambda^+_c\to p,\;\; & \Sigma^+_c(2455)\to p, & \Sigma^0_c(2455)\to n, \\
\Xi^+_c\to\Sigma^+, & \Xi^0_c\to\Sigma^0, & \Omega^0_c\to\Xi^0,
\end{eqnarray}
with the amplitudes given by the expression (\ref{MatrixElement}).
Due to the significant difference between masses of the initial and final baryon states, we cannot use the approximation for the form factors $f_i(0)\approx 0$, $g_i(0)\approx 0$, $i=1,2$. We then use the values obtained in Ref.\cite{Luo:1998wg} under the condition that the decay of the baryon is carried by the ``heavy quark''$\rightarrow$``light quark'' transition.
The coefficients $m_A$, $m_V$ for the charmed baryons in the dipole approximation (\ref{DipoleApprox}) of $f_1$, $g_1$ are equal to
 \cite{Cheng:1995fe}
\begin{eqnarray*}
m_V=2.11 \text{GeV}, & \quad m_A=2.54 \text{GeV}.
\end{eqnarray*}
We perform calculations of the squared amplitudes with the help of FeynCalc package \cite{Mertig:1990an}.
The resulting expression consists of 1042 terms, so we do not quote it here. Finally, we perform numerical integration 
over the phase space. Results for the decays listed in Eq. \eqref{charmed reactions} are summarized in the Table \ref{Table}.

\begin{table}[h!]
 \caption{Decay widths and branching ratios. $\Gamma_\text{W}$, $\Gamma_{\text{em,W}}$ and $ \Gamma_{\text{em}} $ are defined in Eq. \eqref{SigmaPWidth} }
 \label{Table}
\begin{center}
\begin{tabular}{|c|c|c|c|c|}
\hline
Decay & $\Gamma_\text{W} \text{,GeV}^5$ & $\Gamma_{\text{W,em}} \text{,GeV}^5$ & $ \Gamma_{\text{em}} \text{,GeV}^5$ & 
$Br\vert_{R<R_{K\to\mu e}}$\\
\hline
$\Lambda^+_c\to p\mu e^+$ & $7.7\cdot10^{-4}$ & $7.2\cdot10^{-5}$& $5.3\cdot10^{-5}$ & $4.6\cdot 10^{-14}$\\
\hline
$\Sigma^+_c\to p\mu e^+$& $5.8\cdot10^{-6}$ & $2.5\cdot10^{-6}$& $8.7\cdot10^{-6}$ & $4.4\cdot 10^{-25}$\\
\hline
$\Sigma^0_c\to n\mu e^+$& $1.4\cdot10^{-5}$ & $4.4\cdot10^{-6}$& $2.0\cdot10^{-5}$ & $2.0\cdot 10^{-24}$\\
\hline
$\Xi^+_c\to\Sigma^+\mu e^+$& $3.7\cdot10^{-5}$ & $7.7\cdot10^{-6}$& $1.1\cdot10^{-5}$ & $5.7\cdot 10^{-15}$\\
\hline
$\Xi^0_c\to\Sigma^0\mu e^+$& $1.8\cdot10^{-5}$ & $3.8\cdot10^{-6}$& $5.7\cdot10^{-6}$ & $2.8\cdot 10^{-15}$\\
\hline
$\Omega^0_c\to\Xi^0\mu e^+$& $1.0\cdot10^{-6}$ & $1.1\cdot10^{-5}$& $3.0\cdot10^{-5}$ & $5.8\cdot 10^{-17}$\\
\hline
$\Lambda^0_b\to\Sigma^0\mu e^+$& $9.3\cdot10^{-4}$ & $3.0\cdot10^{-3}$& $6.7\cdot10^{-3}$ & $2.34\cdot 10^{-12}$\\
\hline
$\Lambda^0_b\to n\tau e^+$& $1.0\cdot10^{-5}$ & $3.3\cdot10^{-5}$& $7.6\cdot10^{-5}$ & $2.6\cdot 10^{-14}$\\
\hline
$\Xi^0_b\to\Xi^0\mu e^+$& $1.0\cdot10^{-3}$ & $3.3\cdot10^{-3}$& $7.4\cdot10^{-3}$ & $2.7\cdot 10^{-12}$\\
\hline
$\Xi^0_b\to\Sigma^0\tau e^+$& $5.1\cdot10^{-4}$ & $1.6\cdot10^{-3}$& $3.7\cdot10^{-3}$ & $1.3\cdot 10^{-12}$\\
\hline
$\Xi^-_b\to\Xi^-\mu e^+$& $5.7\cdot10^{-3}$ & $7.2\cdot10^{-3}$& $1.6\cdot10^{-2}$ & $7.3\cdot 10^{-12}$\\
\hline
$\Omega^-_b\to\Omega^-\mu e^+$& $6.8\cdot10^{-3}$ & $2.2\cdot10^{-2}$& $5.1\cdot10^{-2}$ & $1.2\cdot 10^{-11}$\\
\hline
$\Omega^-_b\to\Xi^-\tau e^+$& $8.1\cdot10^{-3}$ & $2.7\cdot10^{-2}$& $6.2\cdot10^{-2}$ & $1.5\cdot 10^{-11}$\\
\hline
\end{tabular}
\end{center}
\end{table}

\subsubsection*{Beauty baryons}
Forbidden decays of beauty baryons occur by $b\to d$ and $b\to s$ transitions. Their consideration repeats the previous discussion. We are interested in $e\to\tau$ lepton current transition for $b\to d$ reaction, and $e\to\mu$ transition for $b\to s$ one, which are $\vert \Delta N \vert =0 $. The appropriate parts of the effective Lagrangian are similar to (\ref{First_vertex}), (\ref{SecondVertex}) with the obvious replacements. The constant $\zeta$ in the expressions for the coupling constants \eqref{G} turns to \cite{Libanov:2012kh}
\begin{equation*}
\zeta'=\sum_{l=1}^{\infty}\dfrac{\left(E^{l,2}_{13}\right)^2}{l(l+1)}\approx 0.27
\end{equation*}
for $b\to d$, $e\to\tau$ transitions and
\begin{equation*}
\zeta''=\sum_{l=1}^{\infty}\dfrac{E^{l,1}_{12}E^{l,1}_{23}}{l(l+1)}\approx 0.47
\end{equation*}
for $b\to s$, $e\to\mu$ transitions.
For the coefficients in the dipole approximation (\ref{DipoleApprox}) of $f_1$, $g_1$ we have  \cite{Cheng:1995fe}:
\begin{equation*}
\begin{array}{cc}
m_V=6.34 \text{GeV}, & m_A=6.73 \text{GeV}.
\end{array}
\end{equation*}
The list of the processes and the corresponding decay widths is presented in the Table \ref{Table}.

\section{conclusions}
We have investigated the semileptonic baryon decays provided by the currents that are forbidden within the SMб but that, however, can take place in its certain extensions.
Since the processes changing the total generation number $\Delta N$ are suppressed 
in the model, we considered only those with $\Delta N = 0$. The restrictions on $1/R$, calculated in this work, appeared to be weaker than the similar restrictions obtained from $B$- and, especially, $K$-meson decays \cite{Frere:2003ye}.
 The current experiments, including LHCb, could not provide sufficient statistics to increase
  the low limit on $R^{-1}$ up to the level $\sim 10TeV$.  

\section{ACKNOWLEDGMENTS}
The authors are indebted to D. S. Gorbunov, M. V. Libanov, E. Ya. Nugaev and K.O. Astapov for helpful discussions. This work was supported by RFBR Grant No. 14-02-31813.

\bibliography{refs}

\end{document}